\documentclass[twocolumn,prl,showpacs]{revtex4}
\usepackage{graphicx}
\usepackage{dcolumn}
\usepackage{bm}
\usepackage{amsmath}

\begin{document}

\preprint{}
\title{Gate-controlled one-dimensional channel on the topological surface}
\author{Takehito Yokoyama$^{1}$, Alexander V. Balatsky$^{2}$, and Naoto Nagaosa$^{1,3}$}
\affiliation{
$^1$Department of Applied Physics, University of Tokyo, Tokyo 113-8656, Japan \\
$^2$Theoretical Division and Center for Integrated Nanotechnologies, Los
Alamos National Laboratory, Los Alamos, New Mexico 87545, USA \\ 
$^3$Cross Correlated Materials Research Group (CMRG), ASI, RIKEN, WAKO 351-0198, Japan 
}
\date{\today}

\begin{abstract}
We investigate the formation of the one-dimensional channels  on the topological surface under the gate electrode. The energy dispersion of these channels is almost linear in the momentum with the velocity sensitively depending on the strength of the gate voltage. The energy is also restricted to be positive or negative depending on the strength of the gate voltage. 
Consequently, the local density of states near the gated region has an asymmetric structure with respect to zero energy. 
In the presence of the electron-electron interaction, the correlation effect can be tuned by the gate voltage. We also suggest a tunneling experiment to verify the presence of these bound states. 

\end{abstract}

\pacs{73.43.Nq, 72.25.Dc, 85.75.-d}
\maketitle



%

%




Recent discovery of the two-dimensional (2D) quantum spin Hall system\cite{Mele,Bernevig,wu2006,xu2006,Fu,Qi,Bernevig2,Konig},
and its three-dimensional (3D) generalization, dubbed topological insulator, \cite{Moore,Fu2,Teo,Hsieh,Qi} has established a new state of matter 
in the time-reversal symmetric systems.
The topological order in the bulk with the gap guarantees the presence of the one-dimensional (1D) channels along the edge of the 2D sample, or the 2D metal on the surface 
of the 3D sample. These edge and surface states are 
protected by the time-reversal symmetry and the topology of the bulk gap, 
and are robust against the disorder scattering and electron-electron interactions.
   
In topological surface state on 3D topological insulator, 
the electrons obey the 2D Dirac equations. 
This has been beautifully demonstrated by the 
spin- and angle-resolved photoemission spectroscopy~\cite{Hasan,Ando}.
To date, various interesting properties of the topological insulator have been predicted, particularly those relevant to magnetism~\cite{Maciejko,Qi2,Liu,Yokoyama,Tanaka} and electron-electron interaction\cite{Hou,Strom,Tanaka2,Teo2,Raghu,Zhang}.
Since the surface state of topological insulator has a suppressed backward scattering for nonmagnetic impurities, this is a promising material for designing the quantum curcuit. Therefore,  electric control of transport on the surface of topological insulator is an important issue. 
In fact, the chiral edge channels are predicted to appear at the interface of two ferromagnets with magnetization along $z$ and $-z$  directions, i.e., perpendicular to the surface.\cite{Niemi}  Also, scanning tunneling microscopy experiments and related theories have revealed the electronic states due to the impurity scattering or the terrace edge on the surface of topological insulator\cite{Alpichshev,XZhang,Lee,Wang,Biswas,Biswas2}.


In this paper, we study the 1D states that are formed  on the topological surface under the gate electrode. The energy dispersion of these states  is almost linear in the momentum with the velocity sensitively depending on the strength of the gate voltage. The energy is also restricted to be positive or negative depending on the strength of the gate voltage. 
Consequently, the local density of states near the gated region has an asymmetric structure with respect to zero energy. 
With the electron-electron interaction, the correlation effect can be tuned by the gate voltage. We also suggest a tunneling experiment to verify the presence of these bound states.

\begin{figure}[htb]
\begin{center}
\scalebox{0.8}{
\includegraphics[width=8.0cm,clip]{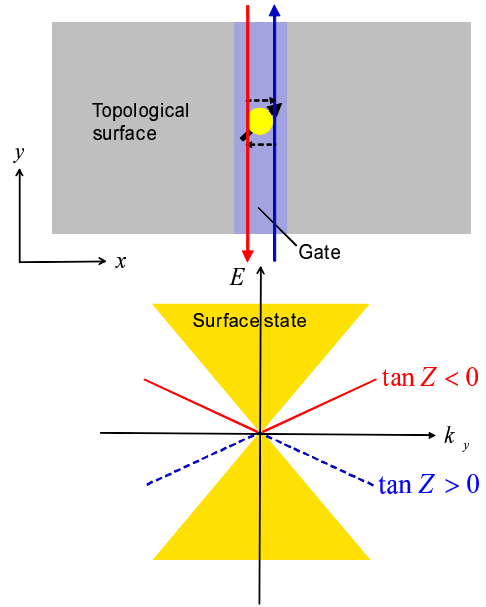}
}
\end{center}
\caption{(Color online) (Upper) schematics of the proposed setup. The gate electrode is attached on the surface of the topological insulator where 1D helical mode is generated. A magnetic impurity placed on the gated region produces the backward scattering. (Lower) dispersions of surface and bound states. 
}
\label{fig1}
\end{figure}

\begin{figure}[htb]
\begin{center}
\scalebox{0.8}{
\includegraphics[width=8.0cm,clip]{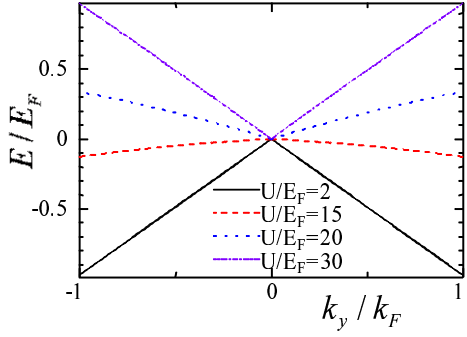}}
\end{center}
\caption{(Color online)  The dispersion of the bound states as a function of $k_y$ for various $U/E_F$ with $k_F L =0.1$ and $E_F  = v_F k_F$. }
\label{fig3}
\end{figure}

We consider 2D topological surface on which the 1D gate electrode is attached as shown in Fig. \ref{fig1}. The gate electrode is attached in the region $0 \le x \le L$. The interface between gated and ungated regions is parallel to the $y$-axis.
The electrons on the topological surface obey the 2D Dirac Hamiltonian
$H = v_F \bm{k} \cdot \bm{\sigma}$ and under the gate, we add the potential $U$, namely $H  = v_F \bm{k}' \cdot \bm{\sigma}  + U$.\cite{note} Here, $v_F$ is the Fermi velocity, $\bm{k} (\bm{k'})$ is a wavevector and $\bm{\sigma}$ is a vector of Pauli matrices.  
The dispersion relation is $E =  \pm v_F \sqrt {k_x^2  + k_y^2 } =  \pm v_F \sqrt {k_x^{'2}  + k_y^2 }  + U.$ Note that due to the translational symmetry along the $y$-axis, $k_y$ is conserved.
Now, let us regard $v_F \left| {k_y^{} } \right|$ as a gap for a fixed $k_y$ and consider the bound states formed inside this gap.
By setting $k_x  = i\kappa (\kappa>0)$ outside the gated region to search for the bound states near the gate, with the above Hamiltonian, 
we have wave functions in each region represented as  
\begin{eqnarray}
 \psi (x \le 0) = ae^{\kappa x} \left( {\begin{array}{*{20}c}
   E  \\
   {iv_F (k_y^{}  - \kappa )}  \\
\end{array}} \right), \\ 
 \psi (0 \le x \le L) = be^{ik_x 'x} \left( {\begin{array}{*{20}c}
   {E - U}  \\
   {v_F (k_x ' + ik_y^{} )}  \\
\end{array}} \right) \nonumber \\ + ce^{ - ik_x 'x} \left( {\begin{array}{*{20}c}
   {E - U}  \\
   {v_F ( - k_x ' + ik_y^{} )}  \\
\end{array}} \right), \\ 
 \psi (x \ge L) = de^{ - \kappa x} \left( {\begin{array}{*{20}c}
   E  \\
   {iv_F (k_y^{}  + \kappa )}  \\
\end{array}} \right).
\end{eqnarray}
In the above expressions, the common factor $e^{ik_y y} $ is omitted for simplicity. 

By matching the wavefunctions at the interfaces $x=0$ and $L$, we obtain 
\begin{eqnarray}
v_F^2 k_y^2  + v_F^2 \kappa \frac{{k_x '}}{{\tan k_x 'L}} = E(E - U) 
\end{eqnarray}
where $v_F k_x ' = \sqrt {(E - U)^2  - v_F^2 k_y^2 }$.

By solving the above equation numerically, we plot the dispersion of the bound states as a function of $k_y$ for various $U/E_F$ with $k_F L =0.1$ and $E_F  = v_F k_F$ in Fig. \ref{fig3}. As seen, the bound states are generated with an almost linear dispersion in the momentum. The sign and the velocity of this dispersion strongly depend on the potential parameter $U$.

In order to make the following discussion more transparent, let us take the limit $U \to \infty$ and $L \to 0$ while keeping $Z \equiv UL = const$. Then, we have $v_F \kappa  =  - E\tan Z (> 0)$, and hence $E =  \pm v_F (\cos Z)k_y$ and $\kappa  =  \mp (\sin Z)k_y$. 
When the sign of $\tan Z$ is fixed, then that of $E$ is determined. There are two branches of the bound states, which constitute a 1D helical mode. This is a physical realization of the Tomonaga model which is characterized by two dispersions defined for positive energy only (and no dispersion for negative energy).\cite{Tomonaga}

How does this helical mode manifest itself in observable quantities? In the following, we show its salient feature in local density of states and the current along the 1D channel.
First, we study the local density of states near the gated region ($x \le 0$).
The wavefunction of the bound states is given by 
\begin{eqnarray}
\psi _B (x \le 0) = \sqrt {\frac{\kappa }{{1 \pm \sin Z}}} e^{\kappa x} \left( {\begin{array}{*{20}c}
   {\cos Z}  \\
   {i( \pm 1 + \sin Z)}  \\
\end{array}} \right).
\end{eqnarray}
Note that the spins of the helical mode are opposite to each other. 
Namely, the expectation value of the Pauli matrices is given by 
$\left\langle \bm{\sigma}  \right\rangle \parallel \left( {0, \pm \cos Z, \mp \sin Z} \right)^t$.

The total local density of states can be calculated as 
\begin{eqnarray}
\rho (E ,x) = \frac{1}{{4\pi ^2 }}\int {dk_x dk_y \left| {\psi _S^{} } \right|^2 \delta (E  - E_S )}  \nonumber \\ + \frac{1}{{2\pi }}\int {dk_y \left| {\psi _B^{} } \right|^2 } \delta (E  - E_B )\\
  = \rho ^{(0)}  + \delta \rho (E ,x) + \rho _B (E ,x)
\end{eqnarray}
with $\rho ^{(0)}  = \frac{{\left| E  \right|}}{{2 \pi v_F^2 }}$ and
\begin{widetext}
\begin{eqnarray}
\delta \rho (E ,x) =  - \frac{E }{{ \pi ^2 v_F^2 }}\int_0^{\pi /2} {d\theta \sin ^2 \theta {\mathop{\rm Im}\nolimits} \left( {e^{ - 2i(E \cos \theta ) x/v_F} \frac{{\sin Z}}{{\cos \theta \cos Z - i{\mathop{\rm sgn}} (E )\sin Z}}} \right)}
\end{eqnarray}
\end{widetext}
where $\psi _S$ is the wavefunction for the scattering state, and $E_S$ and $E_B$ are the energy of the scattering state and the bound states, respectively. $\rho ^{(0)}$ and $\delta \rho$ come from the scattering states. \cite{Biswas2}

The local density of states stemming from the bound states $\rho _B$ is given by\begin{eqnarray}
\rho _B (E ,x \le 0) =  - \frac{E }{ \pi v_F^2 }\frac{{\tan Z}}{{\left| {\cos Z} \right|}}e^{ - 2(E \tan Z) x/v_F} \label{dosb}
\end{eqnarray}
for $E \tan Z <0$ and $\rho _B =0$ for $E \tan Z >0$.
The $\rho ^{(0)}$ and $\delta \rho (E ,x)$ are obtained in Ref. \onlinecite{Biswas2}, while $\rho _B (E ,x)$ has been missed in the literature.

\begin{figure}[htb]
\begin{center}
\scalebox{0.8}{
\includegraphics[width=11.0cm,clip]{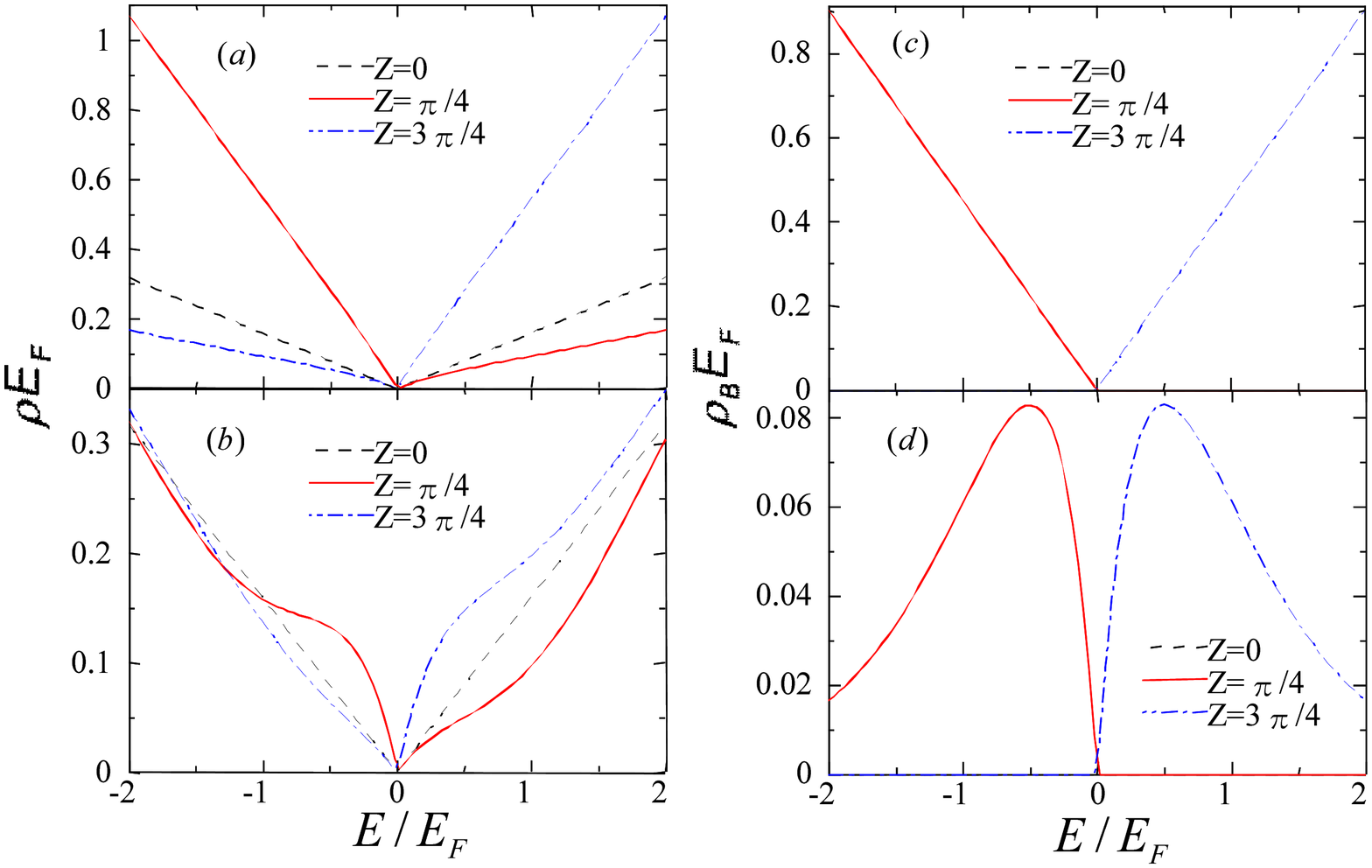}}
\end{center}
\caption{ (Color online) (a) $\rho (E, 0) E_F$, (b) $\rho (E, -1/k_F) E_F$,
(c) $\rho _B (E ,0) E_F$ and (d) $\rho _B (E, -1/k_F) E_F$  for several $Z$. Note $\rho _B (E ,x)=0$ at $Z=0$ and also that the scales of (a), (c) and (b), (d) are different, and the contribution from $\rho_B$ is larger in (a) (at $x=0$) compared with (b) (at $x=-1/k_F$). }
\label{fig4}
\end{figure}

In Fig. \ref{fig4}, we show (a) $\rho (E, 0) E_F$, (b) $\rho (E, -1/k_F) E_F$, (c) $\rho _B (E ,0) E_F$ and (d) $\rho _B (E, -1/k_F) E_F$ for several $Z$. As shown in Fig. \ref{fig4} (a) and Fig. \ref{fig4} (b), the contribution from the bound states gives an asymmetry in the local density of states: $\rho_A(E,x) \equiv \rho(E,x)-\rho(-E,x) \ne 0$ (note $\rho _B (E ,x)=0$ at $Z=0$ and also that $\rho ^{(0)} $ and $\delta \rho$ are even function of $E$).  
 Since the bound states decay exponentially in space as seen in Eq.(\ref{dosb}), the contribution from the bound states becomes strongly suppressed and hence $\rho_A(E,x)$ goes to zero  away from the gated region (compare Fig. \ref{fig4} (c) and Fig. \ref{fig4} (d)). The decay length of the bound states increases with the decrease of ${\left| E  \right|}$, and therefore $\rho _B$ has a peak at low energy as shown in Fig. \ref{fig4} (d), which is reflected in the total density of states  as shown in Fig. \ref{fig4} (b). 
Therefore, by comparing  $\rho_A(E,x)$ near and away from the gated region using scanning tunneling microscopy, one can identify the contribution from the  bound states by $\Delta \rho_A(E)=\rho_A(E,0) - \rho_A(E,-\infty)$. 

Next, let us consider the tunneling current along the 1D channel, taking into account the electron-electron interaction. 
Away from the half-filling, the possible scattering processes are dispersive  ($g_d$) and forward ($g_f$) scatterings.\cite{wu2006,xu2006} Near the Fermi
points, the corresponding interactions can be easily taken into account by the standard bosonization method.\cite{Solyom,Giamarchi,Senechal} The bosonized Hamiltonian is 
\begin{eqnarray}
H = \frac{v}{2}\int {dx\left[ {\frac{1}{K}(\partial _x \phi )^2  + K(\partial _x \theta )^2 } \right]}, \\ 
v = \sqrt {\left( {v_F \left| {\cos Z} \right| + \frac{{g_f }}{{2\pi }}} \right)^2  - \left( {\frac{{g_d }}{\pi }} \right)^2 }, \\
K = \sqrt {\frac{{2\pi v_F \left| {\cos Z} \right| + g_f  - 2g_d }}{{2\pi v_F \left| {\cos Z} \right| + g_f  + 2g_d }}} \label{K}
\end{eqnarray}
with $\phi  = \phi _{R \uparrow }^{}  + \phi _{L \downarrow }^{}$ and $\theta  = \phi _{R \uparrow }^{}  - \phi _{L \downarrow }$ where $\phi _{R \uparrow }$ and $\phi _{L \downarrow }$ define chiral boson fields of spin up right mover and spin down left mover, respectively. 
As seen, the Luttinger parameter $K$ is tunable by changing $Z$, namely the gate voltage. This opens up a possilbility of controlling the property of the 1D interacting fermions by gate voltage. 

We consider a single magnetic impurity placed on the gated region, which produces the backward scattering. Then, the tunneling current $I_C$ can be calculated within the linear response regime and is given by  \cite{Strom,Kane,Furusaki}
\begin{eqnarray}
I_C  \sim \left( {eV} \right)^{\frac{2}{K} - 1}
\end{eqnarray}
with an applied voltage $V$. Here, we consider a weak impurity scattering. For a strong impurity scattering, the result will be modified due to formation of the localized impurity resonance.\cite{XZhang,Biswas}
Thus, the dependence of $K$ on $Z$ can be experimentally studied through the tunneling current. Since the backward scattering requires spin flip process, when the magnetic moment of the magnetic impurity is parallel to the spin of the bound states, the backward scattering is prohibited. This means that by changing the direction of the magnetic moment by, e.g. an applied magnetic field, the tunneling current can be tuned. 

\begin{figure}[htb]
\begin{center}
\scalebox{0.8}{
\includegraphics[width=6.5cm,clip]{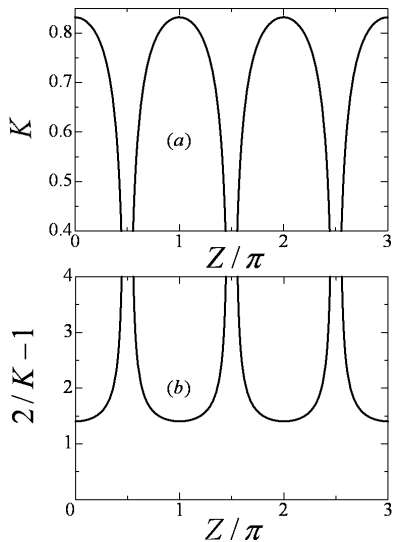}}
\end{center}
\caption{ (a) $K$ and (b) $2/K-1$ as a function of $Z/\pi$ at $g_f/(2\pi v_F)=g_d/(2\pi v_F)=0.1$. }
\label{fig2}
\end{figure}

In Fig. \ref{fig2}, we show (a) $K$ and (b) $2/K-1$ as a function of $Z/\pi$ at $g_f/(2\pi v_F)=g_d/(2\pi v_F)=0.1$.
As shown in Fig. \ref{fig2} (a), the $K$ value oscillates with $Z$ as expected from Eq.(\ref{K}). Near $Z=\pi/2$ mod $\pi$, $K$ is strongly suppressed.
Correspondingly, the exponent of the current, $2/K-1$, has a diverging behavior near 
such points as shown in Fig. \ref{fig2} (b). 
This dependence can be observed in tunneling experiments.

It is noted that the conductance between the source and drain attached to the gated region is still dominated by the 2D surface metal on topological insulator, i.e., $\sigma_{2D} = (e^2/h) k_F \ell >> \sigma_{1D} \sim e^2/h$ with the mean free path $\ell$ and the conductance from the 1D channel $\sigma_{1D}$. One possible way to separate these two contributions is the modulation spectroscopy using the precession of the spin of the magnetic impurity. Namely, the backward scattering potential is dependent on the direction of the spin, which oscillates with the precession, and hence the ac component of the 1D conductance occurs. One catch is that the universal conductance fluctuation (UCF)
of the order of $e^2/h$ also contributes to the ac component of the conductance \cite{UCF}, but the dependence on the spin direction is random, while the 1D channel conductance shows the systematic dependence, i.e., 
it is most suppressed when the backward scattering is maximized. Therefore, by taking the average over
several samples, one can separate the UCF and the 1D conductance of our interest.


In summary, we studied the formation of the one-dimensional channels  on the topological surface under the gate electrode. The energy dispersion of these channels is almost linear in the momentum with the velocity sensitively depending on the strength of the gate voltage. The energy is also restricted to be positive or negative depending on the strength of the gate voltage. 
As a result, the local density of states near the gated region has an asymmetric structure with respect to zero energy. 
With the electron-electron interaction, the correlation effect in the bound states can be tuned by the gate voltage, which can be verified by tunneling experiments.

This work is supported by Grant-in-Aid for Scientific
Research (Grants No. 17071007, 17071005, 19048008 19048015,
and 21244053) from the Ministry of Education, Culture,
Sports, Science and Technology of Japan. T.Y. acknowledges support by JSPS.
Work at Los Alamos was supported by US DOE through LDRD and BES (A.V.B).



\begin{thebibliography}{99}
\bibitem{note} The results can be easily translated to the case of $H  = v_F (\bm{k} \times \bm{e}_z) \cdot \bm{\sigma}$ just by rotating $\bm{k}$ by $\pi/2$.
\end{thebibliography}
\end{document}